\documentclass[conference]{IEEEtran}
\IEEEoverridecommandlockouts

\usepackage{algorithm}
\usepackage{algpseudocode}
\usepackage{amsfonts}
\usepackage{amsmath}
\usepackage{booktabs}
\usepackage{graphicx}
\usepackage{xcolor}
\usepackage{multirow}

\newcommand{\ClusterVLM}{\textsc{Mosaic}}

\begin{document}

\title{\ClusterVLM{}: Cross-Modal Clustering for Efficient Video Understanding}

\author{
\IEEEauthorblockN{Tuowei Wang}
\IEEEauthorblockA{Tsinghua University\\
Beijing, China}
\and
\IEEEauthorblockN{He Zhou}
\IEEEauthorblockA{Tsinghua University\\
Beijing, China}
\and
\IEEEauthorblockN{Chengru Song}
\IEEEauthorblockA{Kuaishou Technology\\
Beijing, China}
\and
\IEEEauthorblockN{Qiushi Li}
\IEEEauthorblockA{Beihang University\\
Beijing, China}
\and
\IEEEauthorblockN{Ju Ren* \thanks{Corresponding author: renju@tsinghua.edu.cn.}}
\IEEEauthorblockA{Tsinghua University\\
Beijing, China}
}

\maketitle 

\begin{abstract}
Large vision-language models (VLMs) are enabling interactive video reasoning, giving rise to \textit{streaming long-video understanding}. In this setting, frames arrive continuously, while the system preserves long-term context and generates responses under strict latency constraints. A central challenge is KVCache management: as video streams grow, KVCache expands rapidly, increasing computation and memory overhead. Existing retrieval-based approaches exploit attention sparsity and offload inactive KVCache from GPU to CPU memory, but their token-level design causes high management overhead and fragmented data movement. We present \ClusterVLM{}, the first cluster-driven VLM inference system for streaming long-video understanding. Our key insight is that VLM KVCache exhibits an implicit \textit{cross-modal clustering} structure: retrieved KV states form groups jointly shaped by visual coherence and semantic relevance. Based on this observation, \ClusterVLM{} uses cross-modal clusters as the basic unit of KVCache organization, maintenance, and retrieval. Evaluations show that \ClusterVLM{} outperforms state-of-the-art baselines, achieving up to $1.38\times$ speedup.
\end{abstract}

\section{Introduction}
The evolution of large language models (LLMs) has fundamentally reshaped natural language processing~\cite{gpt-5,gemini3-pro}. Building upon this foundation, recent advances in multimodal modeling have augmented LLMs with visual encoders, giving rise to large vision-language models (VLMs) capable of jointly reasoning over text and visual inputs~\cite{vit}. These models have enabled a new class of deployment scenarios, including autonomous driving~\cite{drive-gpt4,lm-drive}, embodied agents~\cite{palm-e,open-vla}, interactive assistants~\cite{llava,ferret}, and video question-answering~\cite{video-llava,video-chat}. 

While early VLM studies primarily centered on static image-text pairs, many of these emerging applications operate over continuous video streams. In this paper, we focus on such \textit{video-centric} VLM deployments, where visual inputs arrive as temporally ordered frame sequences. Compared with text-only LLM applications, two defining characteristics are becoming increasingly prominent in practical VLM workloads:

First, video inputs are inherently \textit{long}. Real-world videos often span hundreds or thousands of frames, each encoded into many visual tokens before being fused with textual representations. As the video grows, the effective context length increases rapidly, resulting in a larger key-value cache (KVCache) and higher computational and memory overhead.

Second, these workloads are inherently \textit{streaming}. Unlike offline video analysis, where the full sequence is available upfront, frames in real deployments arrive continuously and must be processed online. As the stream grows, the KVCache must expand to preserve longer context, making it difficult to generate accurate responses under real-time constraints.

We refer to this combined workload as \textbf{\textit{Streaming Long Video Understanding}}, where the system must simultaneously accommodate long temporal contexts and strict real-time constraints. A natural approach to scaling such workloads is to exploit the inherent sparsity of attention. Although attention is formally defined over the full context, empirical evidence suggests that only a small subset of historical tokens dominates the final output for each query. Based on this observation, recent systems~\cite{rekv,live-vlm,stream-mem} selectively retrieve relevant KVCache entries on demand while offloading inactive KVCache to CPU memory, extending context beyond GPU memory limits.

However, existing retrieval-based designs primarily operate \textbf{at token-level granularity}, which introduces new performance bottlenecks. \textit{First}, the overhead of fine-grained decision-making becomes non-negligible. Per-frame relevance scoring, indexing, and selection impose additional computation and bookkeeping costs, particularly under high frame rates and frequent decoding steps. As the system continuously evaluates and updates candidate frames, the cumulative control overhead can offset the theoretical efficiency gains of sparse attention. \textit{Second}, I/O fragmentation emerges as a critical inefficiency. Retrieval frequently involves transferring small, scattered KV segments between CPU and GPU memory. Such fine-grained data movement leads to poor bandwidth utilization and increased latency due to repeated synchronization and transfer overhead. Consequently, the system becomes bottlenecked by memory traffic rather than attention computation itself.

To overcome these limitations, we identify a previously underexplored property of VLM inference: KVCache is often accessed not as isolated frames, but as relatively stable groups of frames. These groups are shaped jointly by visual and semantic structure. This behavior reveals the presence of \textbf{\textit{Cross-Modal KVCache Clustering}}, arising from the interaction between visual representations and language-conditioned attention. This observation motivates a paradigm shift in KVCache management: \textit{replacing token-level retrieval with cluster-level retrieval to better exploit cross-modal structure}, thereby reducing system overhead and improving efficiency.

\begin{figure}[t]
    \centering
    \includegraphics[width=1.0\linewidth]{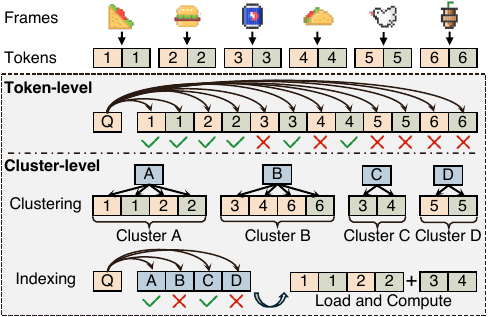}
    \caption{Comparison of token-level and cluster-level KVCache management.}
    \label{fig:comparison}
\end{figure}

In this paper, we propose \textbf{\ClusterVLM{}}, the first clustering-driven VLM inference system for streaming long video understanding. As illustrated in Figure~\ref{fig:comparison}, the key idea of \ClusterVLM{} is to introduce cross-modal clustering as a first-class abstraction for organizing token-level KVCache, enabling efficient KVCache offloading, retrieval, and update. Specifically, \ClusterVLM{} consists of three key components:

\noindent\textbf{Cross-Modal Constructor.} To address the mismatch between existing KVCache management and the actual access patterns of VLMs, \ClusterVLM{} introduces the first cross-modal clustering design for KVCache, which organizes token-level states according to joint visual and semantic structure. Based on this abstraction, \ClusterVLM{} further enables efficient cluster retrieval through a lightweight GPU-resident index.

\noindent\textbf{Self-Adaptive Maintainer.} A static clustering strategy quickly becomes stale in streaming workloads, where new frames continuously reshape the context. \ClusterVLM{} tackles this challenge with a self-adaptive maintenance mechanism that preserves clustering quality online while avoiding the high overhead of frequent global reorganization. This design keeps KVCache clustering effective through long-stream inference.

\noindent\textbf{High-Performance Executor.} To fully realize the efficiency benefits of cluster-based KVCache management, \ClusterVLM{} introduces targeted system optimizations. First, it adopts batched processing to avoid the inefficiencies of fine-grained execution and improve overall hardware utilization. Second, it employs I/O-computation overlap to mitigate the overhead of serialized data movement and computation.

We evaluate \ClusterVLM{} on three popular VLM families and five representative long-video understanding benchmarks across three GPU clusters. Experimental results demonstrate that \ClusterVLM{} outperforms the state-of-the-art solutions in response latency, achieving speedups up to $1.38\times$. Moreover, it reduces GPU memory consumption by up to $2.22\times$, enabling more scalable and efficient long video processing.

We summarize the contribution of this paper as follows:
\begin{itemize}
    \item We identify a new property, cross-modal KVCache clustering, inherent in VLM inference, enabling a paradigm shift in optimizing streaming long video understanding.
    \item We develop three key techniques for organizing, maintaining, and operating on KVCache clusters, encompassing both algorithmic and system-level optimizations.
    \item We implement these techniques as an end-to-end system and comprehensively evaluate its effectiveness across diverse models, benchmarks, and hardware platforms.
\end{itemize}

\section{Background}
\subsection{Large Vision-Language Models}
\begin{figure}[t]
    \centering
    \includegraphics[width=1.0\linewidth]{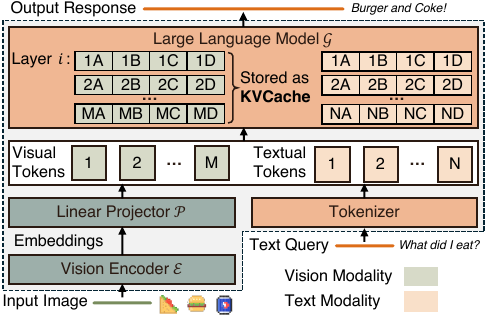}
    \caption{Architecture of the vision-language model.}
    \label{fig:vlm-architecture}
\end{figure}

Large language models (LLMs) have demonstrated remarkable capabilities in natural language understanding and generation by leveraging large-scale transformer decoders. Building upon this foundation, vision-language models (VLMs) extend LLMs to the multimodal setting, enabling joint reasoning over visual inputs and text within a unified architecture. As illustrated in Figure~\ref{fig:vlm-architecture}, modern VLMs typically consists of three components: (1) a \textit{vision encoder} $\mathcal{E}$ that extracts visual embeddings from input images (e.g., ViT-style backbones~\cite{vit}), (2) a \textit{linear projector} $\mathcal{P}$ that maps visual embeddings into the embedding space of language model, and (3) a transformer-based \textit{language model} $\mathcal{G}$ that integrates visual tokens with textual tokens to generate responses token by token.

In video understanding scenarios, a video is first decomposed into a sequence of frames, which are then processed individually by the VLM. These frames are subsequently encoded into a key-value cache (KVCache), so that the visual content of the entire video can be preserved as contextual memory. This design is especially important for long-video applications, as storing precomputed visual representations in the KVCache avoids repeatedly encoding all frames for each query. When a textual query is issued, the corresponding text tokens attend directly to the cached visual tokens, enabling efficient cross-modal reasoning over the video content.

\subsection{Streaming Long Video Understanding}
The adoption of VLMs has enabled more interactive forms of video reasoning. Unlike traditional benchmarks, which operate on short clips that are provided all at once, emerging workloads increasingly involve long-form video streams in which frames arrive continuously. We refer to this new workload paradigm as \textit{Streaming Long Video Understanding}.

\begin{table}[b]
    \centering
    \caption{Response latency and memory footprint of Qwen2.5-VL across different video lengths (in frames).}
    \label{tab:computation-memory-scale}
    \small
    \begin{tabular}{lllllll}
        \toprule
        Video Length & 32f  & 64f  & 128f & 256f & 512f & 1024f \\ \midrule
        Latency      & 2.6s & 4.8s & 9.1s & 16.8s  & 35.1s  & 73.1s   \\
        Memory       & 16GB & 17GB & 18GB & 21GB & 26GB & 36GB  \\
        \bottomrule
    \end{tabular}
\end{table}

\noindent(1) \textit{Long} video inputs impose substantial computational and memory overhead. A single video stream may contain thousands of frames, each encoded into many visual tokens before being fused with textual representations. Consequently, the effective sequence length grows rapidly, resulting in a larger KVCache, which in turn increases attention computation and memory usage. As shown in Table~\ref{tab:computation-memory-scale}, both response latency and memory footprint increase rapidly with video length, limiting the scalability of existing VLM systems to long inputs.

\noindent(2) \textit{Streaming} further complicates this picture. Frames arrive online and cannot be globally reordered or fully preprocessed in advance, while the system must interleave frame ingestion with decoding and produce responses within a fixed latency budget. As the stream grows, the KVCache expands to preserve longer context, making it difficult to generate accurate responses under real-time constraints. Together, these challenges call for a new VLM system design that jointly addresses long-context scalability and streaming efficiency.

\subsection{Retrieval-Based KVCache Offloading}
\begin{figure}[t]
    \centering
    \includegraphics[width=1.0\linewidth]{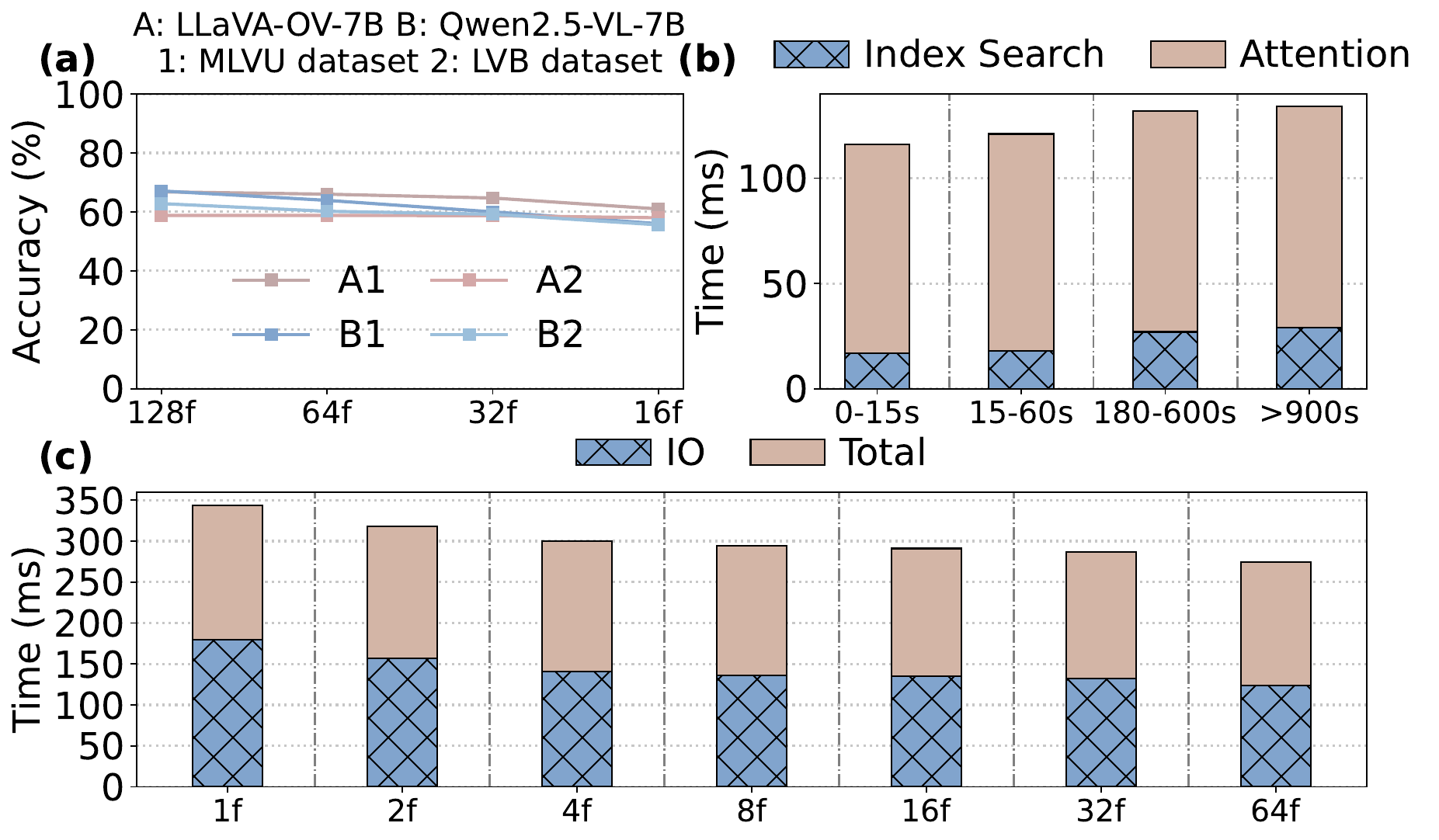}
    \caption{(a) Accuracy under sparse frame sampling (16--128 frames), with only minor degradation as fewer frames are used. (b) Index search time as a percentage of attention time across video durations, ranging from 17\% to 28\%. (c) I/O time decreases with larger block sizes, with a 30\% reduction when increasing from 1 to 64 frames per block.}
    \label{fig:motivation3_abc_combined}
\end{figure}

This combination of unbounded context growth and latency-sensitive inference makes KVCache management a central systems challenge. A key observation behind recent advances is the inherent sparsity of attention: although self-attention is formally defined over the entire sequence, empirical evidence suggests that, for each query, only a small subset of historical tokens receives dominant attention weights, while the majority contributes little to the final output. As shown in Figure~\ref{fig:motivation3_abc_combined}(a), the effective receptive field at each decoding step is often much smaller than the nominal context length, creating opportunities to reduce both computation and memory access. Building on this insight, prior work proposes \textit{KVCache retrieval}, which dynamically selects only the most relevant KVCache for each query instead of attending to the full context. To further extend context capacity, these methods are often combined with \textit{KVCache Offloading}, where inactive KV entries are moved from GPU to CPU memory and fetched back on demand.

In practice, existing KVCache retrieval methods typically operate at token-level granularity, scoring and selecting individual tokens, or their corresponding KV entries, based on relevance. However, this design introduces substantial management overhead. Because retrieval lies on the critical path of decoding, the system must repeatedly evaluate a growing set of candidate frames as the context expands. As shown in Figure~\ref{fig:motivation3_abc_combined}(b), retrieval latency grows with context length and eventually becomes non-negligible, partially offsetting the efficiency gains offered by attention sparsity.

This token-level retrieval also interacts poorly with KVCache offloading. Since the system selectively accesses small pieces of KVCache at fine granularity, it frequently transfers small, scattered KV segments between CPU and GPU. Such fine-grained data movement leads to poor bandwidth utilization and high transfer overhead, resulting in low I/O efficiency. As shown in Figure~\ref{fig:motivation3_abc_combined}(c), end-to-end performance becomes increasingly bottlenecked by memory traffic rather than attention computation, highlighting the need for a more structured and I/O-efficient KVCache management strategy.

\section{Motivation and Challenges}
\subsection{Observation: Cross-Modal KVCache Clustering}
The inefficiency of token-level retrieval suggests streaming long-video inference requires a better KVCache management abstraction. Instead of treating each frame as an independent unit, we seek a higher-level organization that matches how the model actually accesses historical context during decoding.

\begin{figure}[t]
    \centering
    \includegraphics[width=1.0\linewidth]{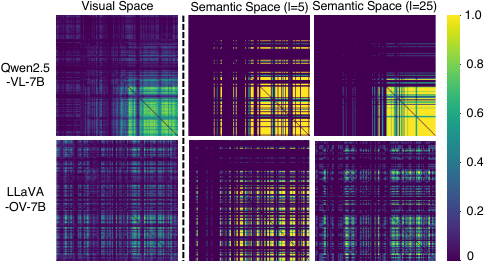}
    \caption{Visualization of KVCache clustering in visual space (left) and semantic space (right) across models and layers on MLVU. Each matrix entry $(i, j)$ represents the frequency with which visual or semantic embeddings $e_{i}$ and $e_{j}$ are retrieved together. Brighter color denotes a larger value.}
    \label{fig:clustering}
\end{figure}

Our key observation is that VLM KVCache exhibits an implicit \textit{cross-modal clustering} structure. Specifically, the KVCache activated by each query is rarely composed of isolated frames; instead, retrieval consistently selects relatively stable groups of frames. As illustrated in Figure~\ref{fig:clustering}(left), the frames activated during retrieval are often visually clustered, forming coherent groups that align with scenes or event segments rather than appearing as isolated instances. At the same time, Figure~\ref{fig:clustering}(right) shows that this structure is also semantic: queries referring to similar entities, actions, or events repeatedly activate the same groups of frames. Together, these results indicate that the effective retrieval unit in VLMs is not an individual frame, but a cross-modal group of frames that is both visually coherent and semantically co-activated.

This distinction is important because neither modality alone is sufficient to capture the observed access pattern. Clustering frames solely by visual embedding similarity may group visually similar content that is not actually accessed together under downstream queries, while clustering solely by semantic relevance may overlook the strong structural regularities induced by visual continuity in videos. 

This observation motivates a new KVCache management strategy that uses cross-modal clusters as the basic unit of organization and retrieval. By elevating the management granularity from frames to clusters, the system can better align retrieval with actual access patterns, while reducing decision overhead and improving data movement efficiency.

\subsection{Main Challenges}
While cross-modal clustering provides a promising direction, incorporating it into a streaming long-video system presents several intertwined challenges that must be tackled:

\noindent\textbf{Challenge \#1: Construction.} The first challenge is how to construct clusters that faithfully capture the cross-modal structure of KVCache while supporting efficient retrieval. Effective cluster partitioning is essential for robust retrieval and accurate decoding, yet cross-modal clustering must simultaneously account for both visual similarity and semantic relevance, making the clustering problem inherently more complex than single-modal grouping. Moreover, the resulting cluster abstraction must align well with the CPU-GPU memory hierarchy for long-video processing. Particularly, clusters should serve as efficient units for retrieval and on-demand transfer, while minimizing GPU memory usage as much as possible.

\noindent\textbf{Challenge \#2: Maintenance.} The second challenge is how to maintain cluster quality dynamically during encoding without introducing excessive overhead. In streaming workloads, new frames continuously arrive and extend the context, causing the underlying clustering structure to evolve over time and deviate from the initial partition. This dynamic behavior requires efficient runtime adaptation to preserve retrieval quality. However, recomputing clusters globally from scratch is impractical under strict latency constraints, while incremental updates accumulate errors and gradually degrade cluster quality.

\noindent\textbf{Challenge \#3: Execution.} The third challenge is how to perform cluster-based KVCache retrieval efficiently on the inference critical path. First, because frames arrive continuously, processing them individually leads to frequent fine-grained computation and data transfer. Such fine-grained execution is inefficient, introducing substantial overhead and reducing effective hardware utilization. Second, existing designs typically execute KVCache selection, transfer, and computation in a sequential manner. Such serialized execution prevents effective overlap and causes data transfer to become a major bottleneck.

\section{Design Overview}
\begin{figure}[t]
    \centering
    \includegraphics[width=1.0\linewidth]{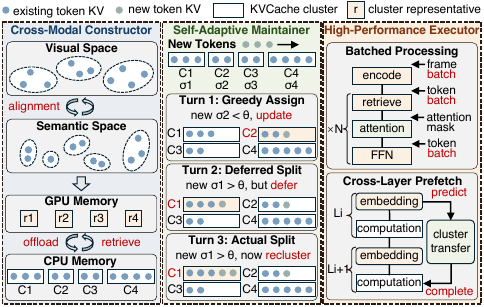}
    \caption{Overview of \ClusterVLM{}.}
    \label{fig:overview}
\end{figure}

To address these challenges, we propose \ClusterVLM{}, the first cluster-driven VLM inference system for streaming long-video understanding. Without requiring any model-level modifications, \ClusterVLM{} adopts a holistic algorithm-system co-design that jointly optimizes KVCache construction, maintenance, and execution. Figure~\ref{fig:overview} presents an overview of \ClusterVLM{}, which consists of three key components.

\noindent\textbf{Cross-Modal Constructor ($\S$~\ref{sec:constructor}).} \ClusterVLM{} first constructs a cross-modal clustering structure over the KVCache that jointly captures visual and semantic relationships. The clustering proceeds in two stages, progressively refining coarse visual groups into semantically aligned clusters. To support efficient retrieval under offloading, \ClusterVLM{} further builds a lightweight GPU-resident index, in which each cluster is summarized by a compact representative vector.

\noindent\textbf{Self-Adaptive Maintainer ($\S$~\ref{sec:maintainer}).} To preserve clustering quality in streaming settings, \ClusterVLM{} dynamically maintains the cluster structure as new frames arrive. It monitors cluster quality online and adaptively updates the partition when the underlying KVCache distribution drifts. To minimize maintenance overhead, \ClusterVLM{} further employs a lazy update mechanism that defers expensive reclustering operations until they are necessary during retrieval.

\noindent\textbf{High-Performance Executor ($\S$~\ref{sec:executor}).} Built on top of the cluster abstraction, \ClusterVLM{} provides an efficient execution engine for cluster-based retrieval on the inference critical path. It batches retrieval requests to amortize fine-grained execution overhead, and designs an I/O-computation pipeline that overlaps cluster selection, data transfer, and attention computation. Together, these optimizations reduce retrieval latency and sustain high streaming inference throughput.

\section{Cross-Modal Constructor}~\label{sec:constructor}
\subsection{Problem Formulation}
\noindent\textbf{Setup.} We consider streaming long-video inference with retrieval-based KVCache offloading on a CPU-GPU heterogeneous memory hierarchy. The workflow consists of two stages: (1) \textit{Frame Encoding.} As video frames arrive continuously, the model incrementally encodes new visual inputs and updates a layer-wise KVCache to preserve historical context for subsequent decoding. To support long-video processing, most historical KVCache is offloaded to CPU memory, while only a small working set remains on the GPU. (2) \textit{Query Retrieval.} When a new query arrives, the system retrieves only the relevant subset of KVCache during attention computation, and generates its response based on the historical video context.

\noindent\textbf{Optimization Objectives.} Our goal is to design an efficient KVCache management system for streaming long-video understanding. In particular, we pursue three objectives:

\noindent(1) \textit{High Retrieval Accuracy.} Retrieved entries should cover the truly relevant historical context as precisely as possible, minimizing redundant transfer while preserving quality.

\noindent(2) \textit{Efficient Data Movement.} Since I/O dominates the performance under a CPU-GPU memory hierarchy, data movement should be minimized while fully utilizing available bandwidth.

\noindent(3) \textit{Low Retrieval Overhead.} Retrieval should compare against as few candidates as possible, thereby reducing both computation cost and control overhead during inference.

\subsection{Nested Visual-Semantic Clustering}
\begin{figure}[t]
    \centering
    \includegraphics[width=1.0\linewidth]{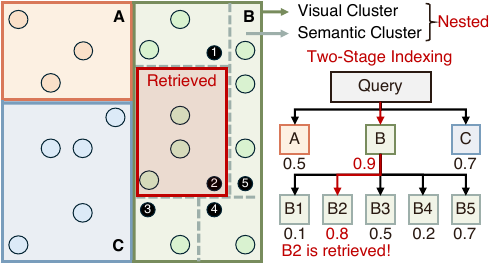}
    \caption{Nested cross-modal clustering and two-stage retrieval indexing.}
    \label{fig:constructor}
\end{figure}

\noindent\textbf{Cross Modality.} Motivated by our observation of cross-modal KVCache clustering, \ClusterVLM{} organizes historical KV states using a nested visual-semantic clustering scheme, as illustrated in Figure~\ref{fig:constructor}. The key idea is to first partition the video stream by visual similarity, and then refine each partition according to layer-specific semantic similarity. This nested design is important for both retrieval accuracy and efficiency.

\noindent(1) \textit{Visual Level.} The first stage clusters frames in the visual embedding space induced by $\mathcal{E}$. These embeddings capture coarse visual regularities such as object appearance, scene layout, and spatial structure. \ClusterVLM{} therefore groups frames with similar visual content into visual clusters, which serve as the top-level partition of the video stream.

\noindent(2) \textit{Semantic Level.} Within each visual cluster, \ClusterVLM{} further clusters KVCache in the semantic space defined by transformer keys. Since KV representations are layer-specific, semantic clustering is performed independently for each layer of $\mathcal{G}$. This second stage refines each visual partition into semantically coherent groups that better match the model’s actual retrieval behavior during decoding.

\noindent\textbf{Clustering Criterion.} In both stages, \ClusterVLM{} adopts a modified K-Means algorithm tailored to the geometry of visual and semantic embeddings. While standard K-Means~\cite{kmeans} relies on Euclidean distance, \ClusterVLM{} instead normalizes embeddings and performs clustering based on cosine similarity. This reduces the effect of raw magnitude differences and better captures directional similarity, which is more informative in the representation space. As a result, the clustering process yields more stable and semantically coherent clusters.

\noindent\textbf{Nested Hierarchy.} A key design in \ClusterVLM{} is to constrain semantic clustering within each visual cluster, rather than performing a global clustering over all KVCache. This nested hierarchy improves retrieval accuracy (\textit{Objective 1}) because it better reflects the cross-modal structure of long-video memory: coarse visual similarity captures stable appearance-level regularities, while fine-grained semantic clustering captures layer-specific retrieval relevance. It also improves system performance (\textit{Objective 2}) by localizing both maintenance and lookup. Newly arrived states only need to be compared against clusters within their corresponding visual partition, and query-time retrieval can first narrow the search space at the visual level before performing finer semantic selection. In this way, \ClusterVLM{} better aligns visual structure with semantic retrieval, making KVCache management consistent with the cross-modal access patterns of VLMs.

\subsection{Hierarchical Indexing and Retrieval}
\noindent\textbf{Cluster Indexing.} All clusters are offloaded to CPU memory and treated as retrieval units that can be fetched on demand. To enable efficient cluster-level retrieval, \ClusterVLM{} adopts an index-based design inspired by database systems. Each cluster is summarized by a representative vector. For a cluster $C_j$, its representative $r_j$ is defined as the centroid:
\begin{equation}
r_{j} = \frac{1}{|C_{j}|}\sum_{k \in C_{j}} k.
\end{equation}
Together, these representative vectors form a compact GPU-resident index that supports efficient cluster lookup.

\noindent\textbf{Two-Stage Retrieval.} Under the nested visual-semantic hierarchy, retrieval is performed in two stages. Given a query, \ClusterVLM{} first compares it against the visual-level representatives to identify the most relevant visual partitions. It then restricts the search to these selected partitions and compares the query with the layer-specific semantic-level representatives to determine the final clusters to retrieve. The selected clusters are then transferred from CPU to GPU for computation. Compared with token-level retrieval, this hierarchical design reduces the number of retrieval candidates (\textit{Objective 3}), increases the granularity of data movement, and better aligns retrieval with the cross-modal structure of KVCache.

\noindent\textbf{Retrieval Augmentation.} To further improve inference quality, \ClusterVLM{} augments the retrieved clusters with two lightweight sources of complementary context. These components complement sparse retrieval by preserving both global historical structure and recent temporal continuity.

\noindent(1) \textit{Global representatives.} The representatives of all clusters are retained as a compact summary of long-range context. Stored in temporal order, they provide a coarse yet global view of the video history, allowing the model to be aware of distant events even when those clusters are not retrieved.

\noindent(2) \textit{Local window.} A small temporal window of recent KV states is maintained to preserve short-range continuity and local dynamics. This is particularly important in streaming video understanding, where adjacent frames are often highly correlated and recent context is relevant to the current query.

\section{Self-Adaptive Maintainer}~\label{sec:maintainer}
\subsection{Cohesion-Aware Adaptive Assignment}
As streaming long-video inference proceeds, new frames and KV states continuously arrive, causing the optimal cluster partition to evolve over time. If the cluster structure is not adjusted appropriately, retrieval quality gradually degrades due to distribution drift. \ClusterVLM{} addresses this challenge with an adaptive online assignment mechanism that tracks cluster cohesion and updates the partition only when necessary.

\noindent\textbf{Cohesion Tracking.} The key challenge of online maintenance is to determine, with minimal overhead, whether newly arrived KV states remain compatible with the current cluster structure. \ClusterVLM{} addresses this problem through a cohesion tracking mechanism based on intra-cluster variance. Since clusters are constructed through K-Means-style partitioning, variance provides a natural measure of cluster compactness:
\begin{equation}
\sigma_{j}^{2} = \frac{1}{|C_{j}|}\sum_{k \in C_{j}} |k - r_{j}|_{2}^{2}.
\end{equation}
A low variance indicates that the elements within a cluster remain semantically coherent, whereas a high variance suggests that the cluster has absorbed increasingly heterogeneous states and may no longer serve as a reliable retrieval unit.

When a new KV state $k_{\text{new}}^{(t)}$ arrives, \ClusterVLM{} greedily assigns it to the cluster whose representative is most similar to the new entry. Let $C_{j}$ denote the selected cluster, with centroid $r_{j}$, variance $\sigma_{j}^{2}$, and size $n_{j}$. After the assignment, both the centroid and the variance are updated online as follows:
\begin{equation}
r_{j}^{\mathrm{new}} = \frac{n_{j} \cdot r_{j} + k_{\text{new}}^{(t)}}{n_{j} + 1}.
\end{equation}
\begin{equation}
(\sigma_{j}^2)^{\mathrm{new}} = \frac{n_{j} \cdot \sigma_{j}^{2} + \left|k_{\text{new}}^{(t)} - r_{j}^{\mathrm{new}}\right|_{2}^{2}}{n_{j} + 1}.
\end{equation}
These updates require only the maintained cluster statistics and do not revisit historical entries, making the mechanism well-suited to streaming settings.

\noindent\textbf{Variance-Guided Handling.} Based on the updated variance, \ClusterVLM{} handles each assignment in a case-by-case manner. If the variance of the updated cluster remains below a threshold, the cluster is regarded as sufficiently cohesive and the new entry is absorbed directly. Otherwise, the cluster is marked as invalid and scheduled for further refinement. In this way, intra-cluster variance serves as the key signal for adaptation, enabling \ClusterVLM{} to detect emerging structural drift without resorting to expensive global reclustering.

Rather than using a fixed value, the threshold is chosen adaptively because cluster geometry varies substantially across transformer layers. We find that early-layer representations are typically more compact, while deeper-layer representations are often more dispersed, so a single threshold is unlikely to work well throughout the model. To address this variation, \ClusterVLM{} uses a size-adaptive threshold:
\begin{equation}
\tau(N)=\tau_{\min}+(\tau_{\max}-\tau_{\min})\exp\left(-\frac{N}{N_{0}}\right),
\end{equation}
where $N$ is the cluster size, $\tau_{\max}$ is the stricter threshold for small clusters, $\tau_{\min}$ is the looser threshold for large clusters, and $N_{0}$ controls the transition rate. The intuition is straightforward: small clusters are often unstable and should not be split aggressively, whereas large clusters are more likely to contain heterogeneous states and should be refined more readily. Accordingly, the threshold relaxes as the cluster grows, allowing refinement to become more robust and adaptive.

\subsection{I/O-Efficient Deferred Splitting}
A straightforward cluster refinement strategy is to split any invalid cluster as soon as its variance exceeds the threshold. In a heterogeneous memory hierarchy, such eager refinement is inefficient: the affected clusters may already have been offloaded to the CPU, and bringing them back to the GPU solely for maintenance would introduce substantial I/O overhead.

To avoid this cost, \ClusterVLM{} adopts a deferred split strategy that decouples structural refinement from immediate data movement, as summarized in Algorithm~\ref{alg:deferred_split}. When a cluster is marked invalid, the system first checks whether its full contents are already resident on the GPU. If so, the split is executed immediately, and the corresponding representatives are updated (Lines 4-6). Otherwise, the split is deferred: the cluster is marked with a lazy-split flag, and the newly arrived KV state is stored in a lightweight GPU-side buffer as an independent singleton cluster for retrieval (Lines 7-10).

This postponement is safe because invalidation is triggered by the insertion of a newly arrived KV state. The resulting inconsistency is therefore localized: it arises from the interaction between the original cluster and the new element, rather than from a global breakdown of the KVCache structure. By keeping the buffered singleton visible to retrieval (Line 10), \ClusterVLM{} preserves its accessibility during subsequent attention computation while avoiding immediate restructuring.

The actual split is performed only when the original invalid cluster is later retrieved and is therefore loaded back to the GPU memory. At that point, the system merges the buffered entries associated with that cluster, executes the deferred split, and updates the states of the refined clusters (Lines 12-17). By aligning maintenance with the natural retrieval procedure, \ClusterVLM{} avoids maintenance-only transfers and turns proactive restructuring into amortized on-demand updates.

\begin{algorithm}[t]
\caption{Deferred Split for Cluster Maintenance}
\label{alg:deferred_split}
\begin{algorithmic}[1]
\Statex \hspace{-\algorithmicindent}\textbf{Require:} new KV state $k_{\mathrm{new}}$, assigned cluster $C$
\Statex \hspace{-\algorithmicindent}\textbf{State:} representative $r$, variance $\sigma^2$, lazy-split flag $\ell$,
\Statex \hspace{0.35cm} GPU buffer $\mathcal{B}$, and representative set $\mathcal{R}$

\Statex \textit{// Insertion Procedure}
\State $(r,\sigma^2) \leftarrow \textsc{Update}(r,\sigma^2, k_{\mathrm{new}})$
\If{$\sigma^2 \le \tau(|C|)$} \Comment{absorb without splitting}
    \State $C \leftarrow C \cup \{k_{\mathrm{new}}\}$
\ElsIf{$C$ is in GPU memory} \Comment{split immediately}
    \State $\{C^{(a)}, C^{(b)}\} \leftarrow \textsc{Split}(C \cup \{k_{\mathrm{new}}\})$
    \State $\{r^{(a)}, r^{(b)}\} \leftarrow \textsc{Rep}(C^{(a)}, C^{(b)})$
\Else \Comment{defer split materialization}
    \State $\ell \leftarrow 1$
    \State $\mathcal{B} \leftarrow \mathcal{B} \cup \{k_{\mathrm{new}}\}$
    \State $\mathcal{R} \leftarrow \mathcal{R} \cup \{\mathcal{B}\}$ \Comment{register as singleton cluster}
\EndIf

\Statex \textit{// Retrieval Procedure}
\If{$\ell = 1$} \Comment{materialize the deferred split}
    \State $C \leftarrow C \cup \mathcal{B}$
    \State $\{C^{(a)}, C^{(b)}\} \leftarrow \textsc{Split}(C)$
    \State $\{r^{(a)}, r^{(b)}\} \leftarrow \textsc{Rep}(C^{(a)}, C^{(b)})$
    \State $\ell \leftarrow 0,\quad \mathcal{R} \leftarrow \mathcal{R} \setminus \{\mathcal{B}\},\quad \mathcal{B} \leftarrow \emptyset$
\EndIf
\end{algorithmic}
\end{algorithm}

\section{High-Performance Executor}~\label{sec:executor}
Beyond cluster construction and maintenance, \ClusterVLM{} further incorporates system-level optimizations for efficient execution. In particular, it introduces two complementary techniques: batched execution to reduce fine-grained overhead during frame encoding, and overlap-aware prefetch to hide data transfer latency during query retrieval.

\subsection{Batch-Oriented Execution}
A naive streaming method processes video frames strictly one at a time. Although such frame-level execution naturally matches the semantics of inference, it is inefficient on modern accelerators. Processing each frame independently leads to a sequence of small and fragmented GPU workloads, frequent kernel launches, repeated synchronization, and low-bandwidth data transfers. In long-stream scenarios, these overheads accumulate over time and can become a major bottleneck.

To improve execution efficiency without breaking streaming semantics, \ClusterVLM{} reorganizes the pipeline based on \emph{temporal dependency}. The key observation is that not all components in a streaming VLM pipeline are equally constrained by time order. In particular, components that are not temporally constrained can be lifted out of the frame-level execution path and processed in batches. \ClusterVLM{} therefore partitions the pipeline into two classes: temporally independent components, which are executed in batches, and temporally dependent components, which remain sequential.

\ClusterVLM{} applies batched execution to three major components. (1) For visual encoding, it batches consecutive frames and processes them jointly with the vision encoder $\mathcal{E}$. (2) For cluster retrieval, it batches token-level matching across frames by organizing token representations into matrices and computing similarities against existing cluster representatives. (3) For the FFN layers in $\mathcal{G}$, it groups token embeddings from multiple frames and processes them jointly through the feed-forward projections. Together, these components transform fine-grained per-frame operations into dense matrix computations that are much better aligned with GPU execution.

For attention computation in $\mathcal{G}$, decoding in a streaming setting appears inherently sequential. To preserve correctness, \ClusterVLM{} enforces temporal constraints through designed attention masking. By restricting each frame to attend only to the causally available context and masking out future information, \ClusterVLM{} enables parallel attention computation across multiple frames without violating semantics.

\subsection{Overlap-Aware Prefetch}
Even with cluster-level retrieval, data movement remains a major source of latency. In a conventional execution, each layer must first compare the current query against all cluster representatives, then identify the relevant clusters, and finally transfer them from CPU to GPU before attention can proceed. When these steps are performed strictly in sequence, data transfer repeatedly falls on the critical path, causing GPU computation to stall on I/O and limiting overall system throughput.

\ClusterVLM{} is motivated by an important observation in transformer decoding: query embeddings from adjacent layers are often highly similar. This phenomenon is largely induced by residual connections~\cite{resnet}, which preserve a substantial portion of the representation across layers. As a result, neighboring layers tend to retrieve highly overlapping sets of clusters. This cross-layer similarity makes it possible to use the query of the current layer as a reliable proxy for predicting which clusters will be needed in the next layer.

\begin{figure*}[t]
    \centering
    \includegraphics[width=1.0\linewidth]{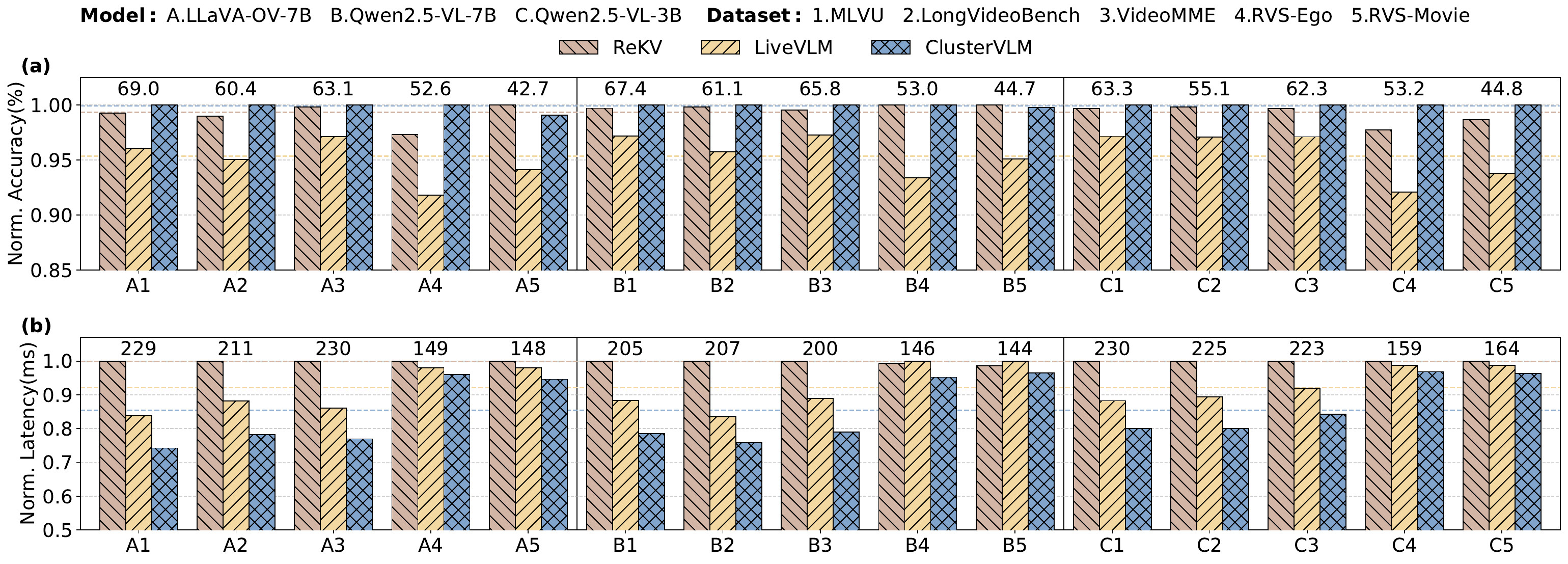}
    \caption{Performance comparison (normalized) between ReKV, LiveVLM and \ClusterVLM{} across models and datasets. Lines indicate average values.}
    \label{fig:exp-overall}
\end{figure*}

\ClusterVLM{} leverages this property to overlap computation with data movement. During layer $l$, the system uses the current query $q^{(l)}$ to predict the top-$K$ clusters that will likely be needed by layer $l+1$ and asynchronously prefetches them to the GPU. At the beginning of layer $l+1$, \ClusterVLM{} uses the actual query $q^{(l+1)}$ to verify whether the prefetched clusters fully cover the retrieval result of the new layer. Any missing clusters are then fetched on demand through a completion step, ensuring correctness before attention computation. By moving most data transfers ahead of time and leaving little recovery work to the next layer, \ClusterVLM{} effectively hides transfer latency while maintaining retrieval quality.

\section{Experimental Evaluation}~\label{sec:evaluation}
\subsection{Experimental Setup}
\noindent\textbf{Hardware.} We conduct experiments on two platforms. Platform A consists of a single NVIDIA H800 GPU (80GB) with an Intel Platinum 8468V CPU. Platform B consists of eight NVIDIA A40 GPUs (48GB each) with an Intel Silver 4314 CPU. Most experiments are conducted on Platform A, while the scalability experiments are conducted on Platform B.

\noindent\textbf{Models.} We evaluate three vision-language models: LLaVA-OneVision-7B~\cite{llava-ov}, Qwen2.5-VL-7B~\cite{qwen25-vl}, and Qwen2.5-VL-3B. These models vary in both architecture and parameter scale, providing strong baselines for video understanding.

\begin{table}[b]
    \centering
    \setlength{\tabcolsep}{2pt}
    \caption{Dataset configurations.}
    \label{tab:datasets}
    \small
    \begin{tabular}{lll}
        \toprule
        \textbf{Dataset}                       & \textbf{Max Length} & \textbf{Description}            \\ \midrule
        MLVU~\cite{mlvu}                       & 703s                & multi-task long video           \\
        \underline{L}ong\underline{V}ideo\underline{B}ench~\cite{long-video-bench} & 468s                & long-term multi-modal video     \\
        VideoMME~\cite{video-mme}              & 1,018s              & full-spectrum multi-modal video \\
        RVS-ego~\cite{rvs}                     & 3,605s              & first-person ego-centric video  \\
        RVS-movie~\cite{rvs}                   & 1,671s              & third-person plot video         \\
        \bottomrule
    \end{tabular}
\end{table}

\noindent\textbf{Dataset.} We evaluate our method on five public video benchmarks, as listed in Table~\ref{tab:datasets}. These benchmarks cover complementary scenarios and video lengths of up to two hours. To further assess effectiveness in real-world deployment, we use a practical streaming long video dataset collected from Kuaishou~\cite{kuaishou}, the largest live-streaming platform in China. In all experiments, videos are processed sequentially to simulate a streaming setting and are sampled at 0.5 FPS.

\noindent\textbf{Baselines.} We compare \ClusterVLM{} against four representative baselines. (1) NoCache uniformly samples frames from the historical video and recomputes attention upon query arrival. (2) ReKV~\cite{rekv} retrieves query-relevant KVCache at the token level. (3) LiveVLM~\cite{live-vlm} further combines token-level retrieval with KVCache compression to reduce memory usage. (4) StreamMem~\cite{stream-mem} also compresses KVCache, but under a query-agnostic memory budget. Following prior work~\cite{rekv,live-vlm}, we set the retrieved frames to 64 for all baselines.

\begin{table}[b]
    \centering
    \setlength{\tabcolsep}{4pt}
    \caption{Performance comparison between NoCache (N), StreamMem (S), and \ClusterVLM{} (C) across models and datasets.}
    \label{tab:exp-overall}
    \small
    \begin{tabular}{llcc}
        \toprule
        \textbf{Metric} & \textbf{Model} & \textbf{MLVU} & \textbf{LongVideoBench} \\
        & & \small (N / S / C) & \small (N / S / C) \\ \midrule
        
        \multirow{3}{*}{Acc. (\%)} 
        & LLaVA-OV-7B   & 66.0 / 66.9 / \textbf{69.0} & 58.6 / 56.4 / \textbf{60.4} \\
        & Qwen2.5-VL-7B & 63.9 / 65.3 / \textbf{67.4} & 60.2 / 57.2 / \textbf{61.1} \\
        & Qwen2.5-VL-3B & 63.8 / 62.3 / \textbf{63.3} & 56.0 / 53.2 / \textbf{55.1} \\ \midrule
        
        \multirow{3}{*}{Lat. (s)}     
        & LLaVA-OV-7B   & 3.26 / 0.16 / \textbf{0.17} & 2.71 / 0.16 / \textbf{0.17} \\
        & Qwen2.5-VL-7B & 3.01 / 0.15 / \textbf{0.16} & 2.69 / 0.15 / \textbf{0.16} \\
        & Qwen2.5-VL-3B & 2.89 / 0.16 / \textbf{0.18} & 2.45 / 0.16 / \textbf{0.18} \\
        \bottomrule
    \end{tabular}
\end{table}

\subsection{Overall Performance}
\noindent\textbf{Accuracy.} As shown in Figure~\ref{fig:exp-overall} and Table~\ref{tab:exp-overall}, \ClusterVLM{} achieves the best accuracy among all methods, outperforming NoCache, ReKV, LiveVLM, and StreamMem by an average of 0.78\%, 0.33\%, 2.53\%, and 2.29\%, respectively, across different models and datasets. NoCache relies on uniform sampling, which often overlooks critical evidence in long videos, while ReKV retrieves frames independently and thus fails to preserve temporal continuity. StreamMem, constrained by a fixed memory budget, tends to lose early contextual information. By contrast, \ClusterVLM{} groups related KVCache through cross-modal clustering, improving quality on both frame encoding and query retrieval.

\noindent\textbf{Retrieval Latency.} We use time-to-first-token (TTFT) as the retrieval latency metric. As shown in Figure~\ref{fig:exp-overall}, \ClusterVLM{} achieves average speedups of $1.28\times$ and $1.12\times$ over ReKV and LiveVLM, respectively, both of which incur substantial overhead from frame-level KVCache management. Table~\ref{tab:exp-overall} further shows that NoCache exhibits much higher latency than \ClusterVLM{} and StreamMem, due to its heavy online attention computation over selected tokens. StreamMem is slightly faster than \ClusterVLM{} because its query-agnostic KV buffer eliminates the need for online retrieval. However, this efficiency comes at the cost of noticeably lower accuracy.

\subsection{Ablation Study}
\noindent\textbf{Nested Clustering.} As shown in Table~\ref{tab:exp-ablation-clustering}, the cross-modal clustering design of \ClusterVLM{} consistently achieves higher accuracy than solely visual clustering, solely semantic retrieval, and global token-level retrieval. A single retrieval modality is often insufficient and leads to suboptimal solutions, whereas global token-level retrieval suffers from substantial noise and redundancy. Instead, \ClusterVLM{} integrates both visual and semantic information with nested clustering.

\begin{table}[b]
    \centering
    \caption{Accuracy comparison across clustering strategies on LLaVA.}
    \label{tab:exp-ablation-clustering}
    \small
    \begin{tabular}{lcc}
        \toprule
        \textbf{Method}     & \textbf{MLVU} & \textbf{LongVideoBench} \\ \midrule
        No Clustering       & 68.4          & 59.8                    \\
        Only Visual Level   & 66.8          & 59.2                    \\
        Only Semantic Level & 68.6          & 59.8                    \\
        \ClusterVLM{}       & \textbf{69.0} & \textbf{60.4}           \\
        \bottomrule
    \end{tabular}
\end{table}

\begin{figure}[t]
    \centering
    \includegraphics[width=1.0\linewidth]{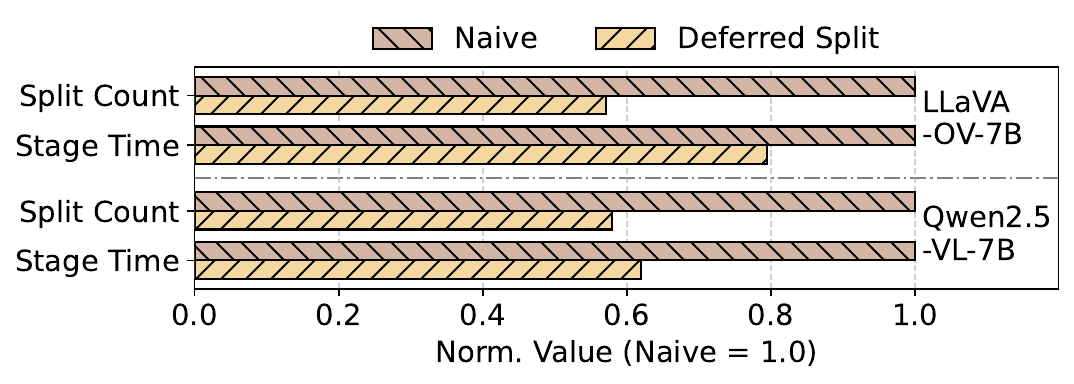}
    \caption{Impact of deferred splitting on maintenance overhead on MLVU.}
    \label{fig:exp-ablation-splitting}
\end{figure}

\noindent\textbf{Deferred Splitting.} As shown in Figure~\ref{fig:exp-ablation-splitting}, deferred splitting reduces the number of cluster split operations during encoding by 42.7\% on average, resulting in a corresponding reduction in latency. In contrast, a naive strategy eagerly splits any invalid cluster as soon as its variance exceeds the threshold, incurring additional I/O overhead when the clusters have been offloaded.

\begin{figure}[t]
    \centering
    \includegraphics[width=1.0\linewidth]{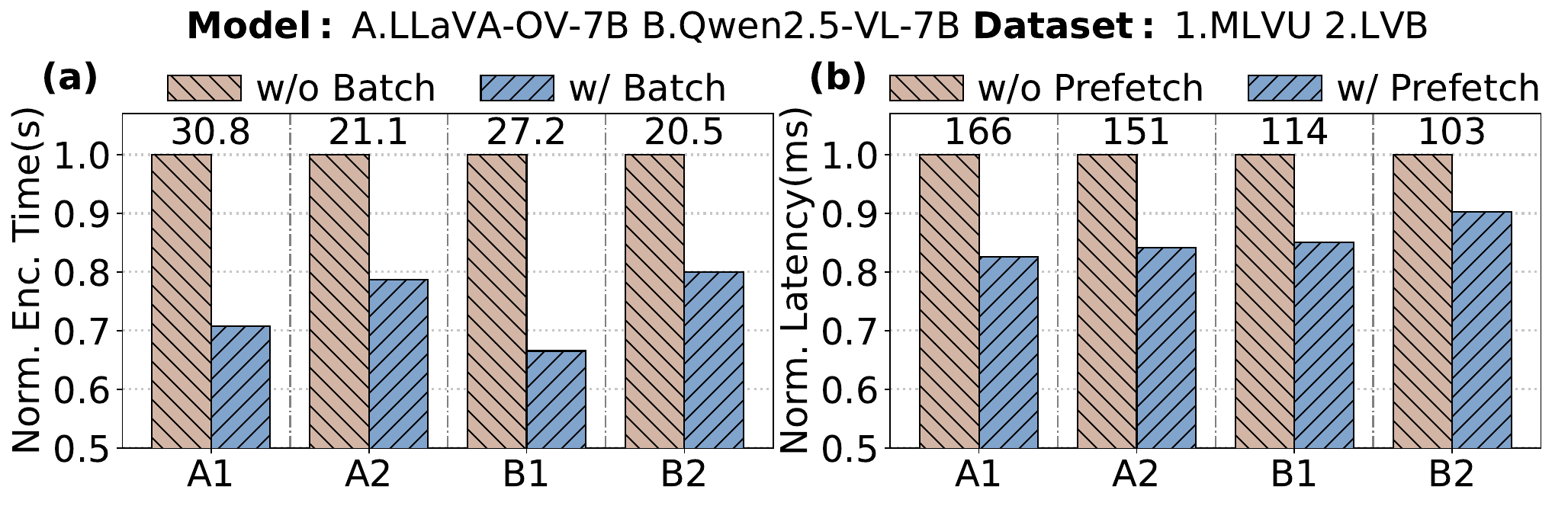}
    \caption{(a) Normalized frame encoding time w/ and w/o batched execution. (b) Normalized response latency w/ and w/o cross-layer prefetch.}
    \label{fig:exp-ablation-system}
\end{figure}

\noindent\textbf{Batched Execution.} Figure~\ref{fig:exp-ablation-system}(a) shows that batched execution reduces frame encoding time by 26\% on average. In contrast, native streaming processes frames one by one, leading to frequent and fragmented operator scheduling. \ClusterVLM{} instead fuses these fragmented operations into dense matrix computations, thereby better exploiting hardware parallelism.

\noindent\textbf{Cross-Layer Prefetch.} Figure~\ref{fig:exp-ablation-system}(b) shows that cross-layer prefetch reduces response latency by 14.5\% on average. This indicates that the prefetch mechanism in \ClusterVLM{} can accurately predict the clusters required by subsequent layers. Therefore, the I/O overhead of sequential retrieval can be effectively overlapped with ongoing computation.

\subsection{Sensitivity Analysis}
\begin{figure}[t]
    \centering
    \includegraphics[width=1.0\linewidth]{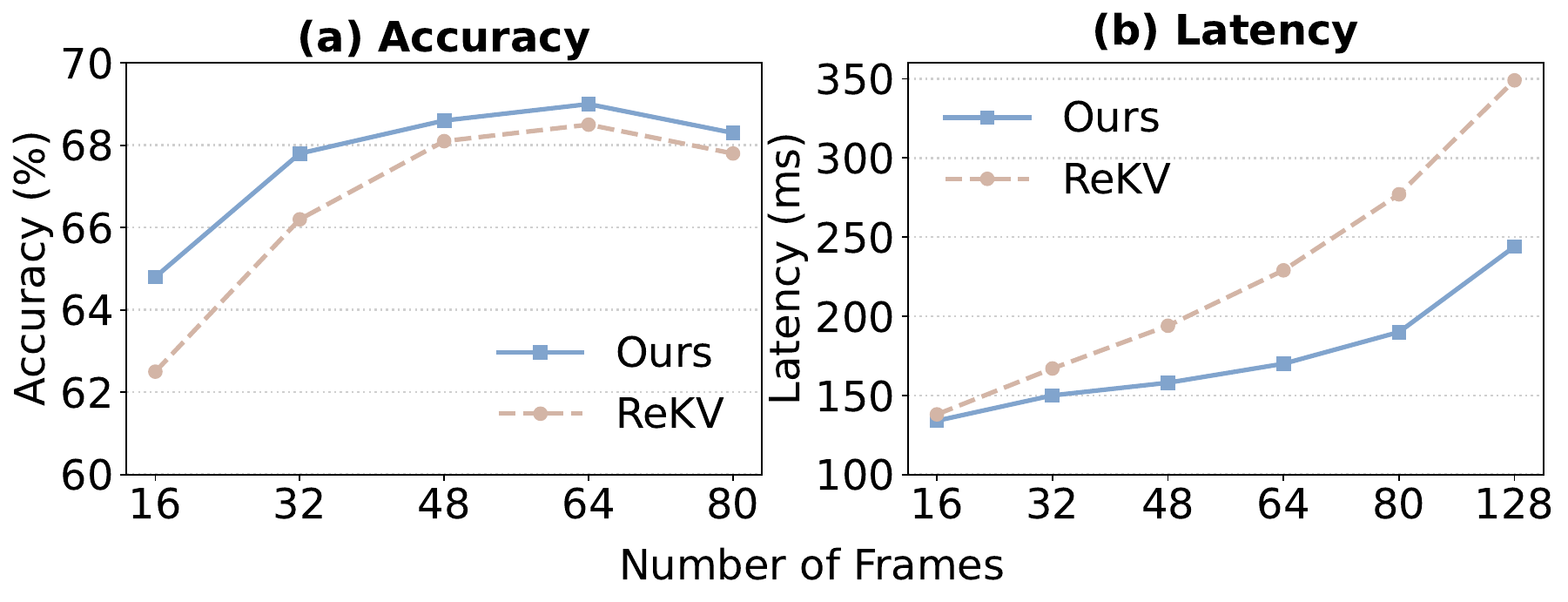}
    \caption{Performance of LLaVA-OV-7B on MLVU across retrieval frames.}
    \label{fig:exp-sensitivity-frames}
\end{figure}

\noindent\textbf{Retrieval Frames.} As shown in Figure~\ref{fig:exp-sensitivity-frames}, \ClusterVLM{} consistently achieves higher accuracy than ReKV across different retrieval frame budgets, with especially pronounced gains under lower budgets. This suggests that \ClusterVLM{} can more effectively identify the most relevant KV caches through cross-modal clustering. In terms of latency, \ClusterVLM{} also consistently outperforms ReKV across all frame budgets, thanks to the higher computational and I/O efficiency of coarse-grained retrieval compared with token-level retrieval.

\begin{figure}[t]
    \centering
    \includegraphics[width=1.0\linewidth]{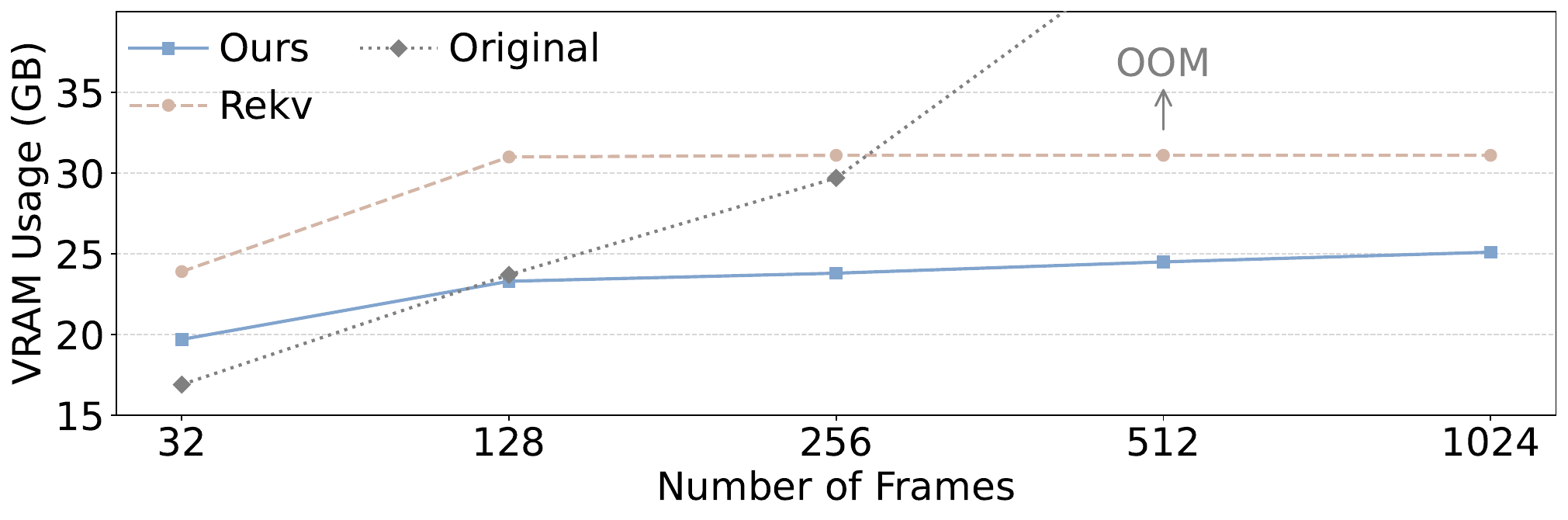}
    \caption{GPU memory usage of LLaVA-OV-7B across different video lengths.}
    \label{fig:exp-sensitivity-vram}
\end{figure}

\noindent\textbf{GPU Memory (VRAM) Usage.} Figure~\ref{fig:exp-sensitivity-vram} shows that when the input video reaches 1024 frames, the unoptimized original model runs out of memory (OOM), while the VRAM usage of ReKV rises to over 30GB. In contrast, \ClusterVLM{} uses only about 25GB VRAM, achieving a 19\% reduction. Conventional token-level retrieval causes VRAM consumption to grow linearly with the number of frames. By managing KVCache through clustering, \ClusterVLM{} needs to store only cluster representatives, thereby reducing VRAM usage.

\begin{figure*}[t]
    \centering
    \includegraphics[width=1.0\linewidth]{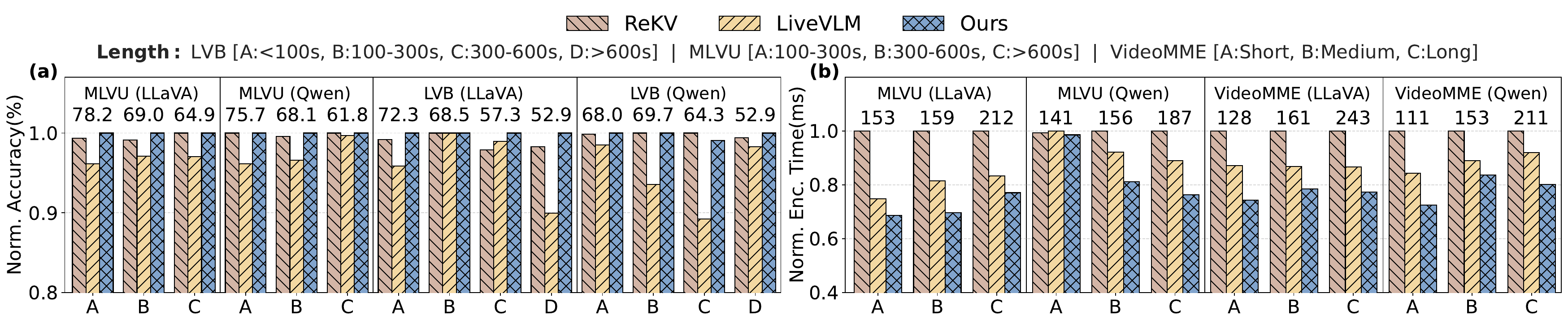}
    \caption{Performance comparison across models and datasets with different video lengths (Qwen denotes Qwen2.5-VL-7B).}
    \label{fig:exp-sensitivity-length}
\end{figure*}

\noindent\textbf{Video Lengths.} Figure~\ref{fig:exp-sensitivity-length} shows that \ClusterVLM{} consistently outperforms all baselines, achieving both the highest average accuracy and the lowest average latency across all video lengths. Through the dynamic maintenance of cross-modal clusters, \ClusterVLM{} performs high-fidelity abstraction over large-scale video streams, enabling efficient real-time question answering on long videos under limited resources.

\begin{figure}[t]
    \centering
    \includegraphics[width=1.0\linewidth]{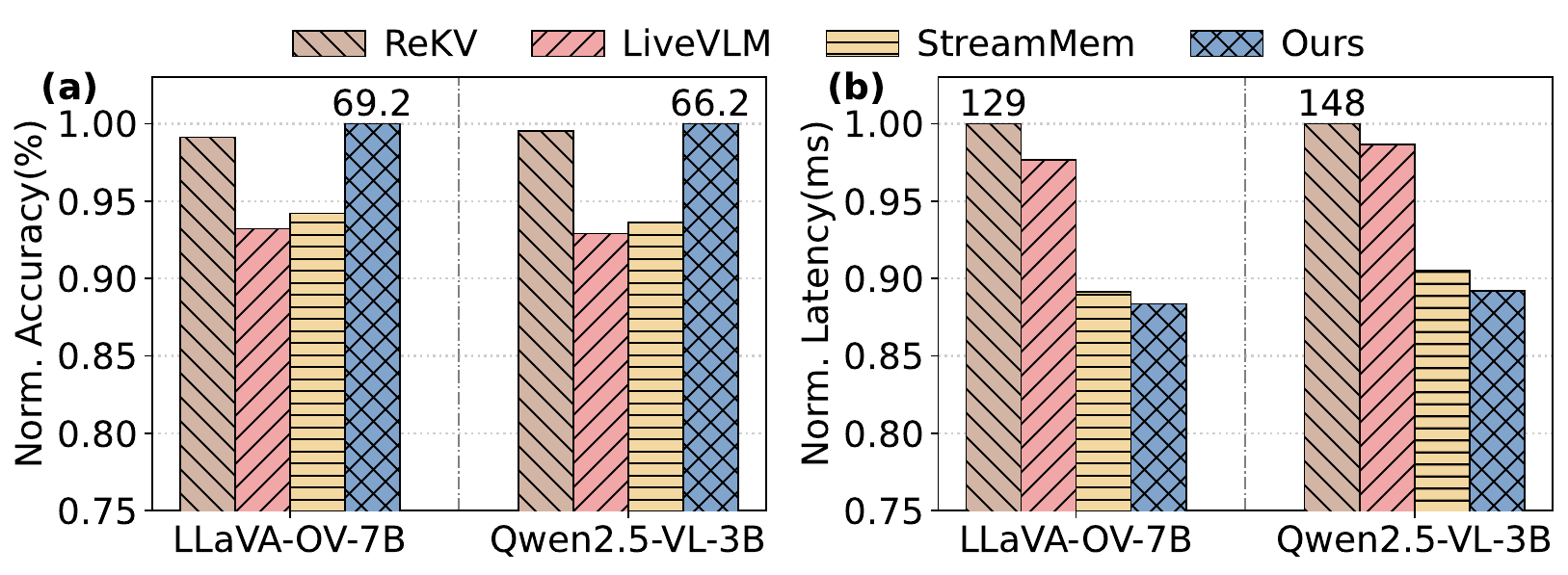}
    \caption{Performance comparison (normalized) on real-world dataset.}
    \label{fig:exp-sensitivity-dataset}
\end{figure}

\noindent\textbf{Real-World Dataset.} We further evaluate \ClusterVLM{} on a real-world dataset collected from Kuaishou live streams. As shown in Figure~\ref{fig:exp-sensitivity-dataset}, \ClusterVLM{} achieves the highest accuracy among all baselines while reducing query latency by 10\%. In these scenarios, frequent scene transitions and redundant static frames are common. \ClusterVLM{} effectively addresses these challenges through its dynamic clustering mechanism, which consolidates redundant static frames efficiently and creates new clusters when scene changes occur.

\begin{figure}[t]
    \centering
    \includegraphics[width=1.0\linewidth]{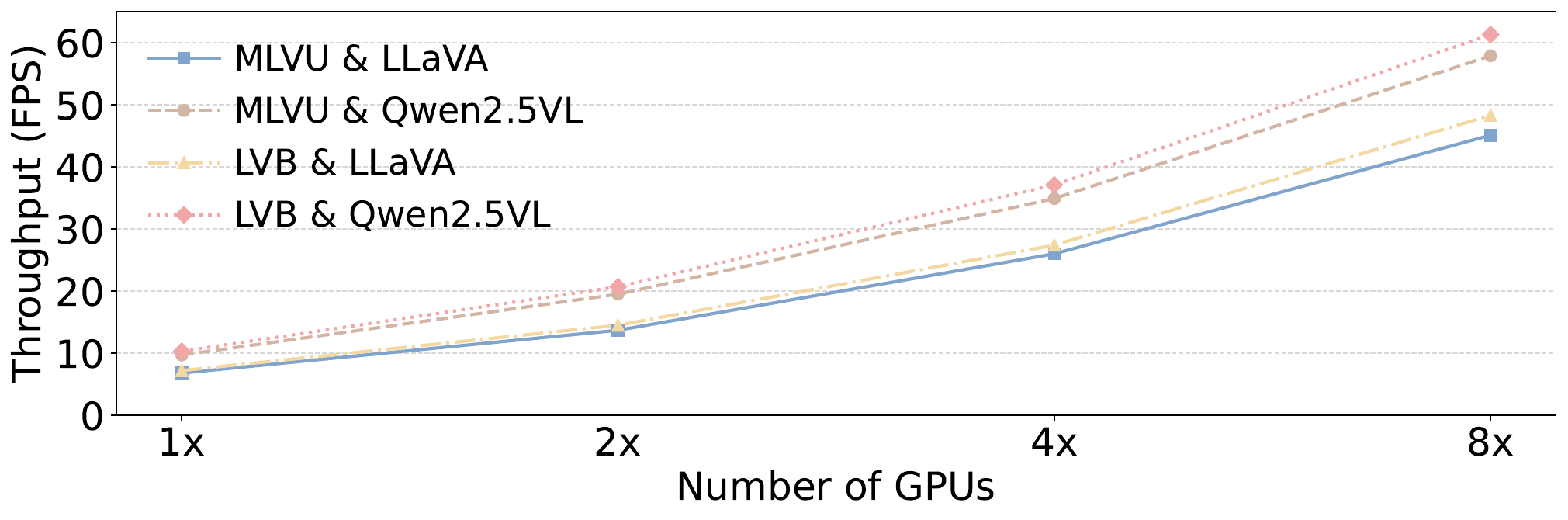}
    \caption{Strong scaling evaluation across different GPU counts.}
    \label{fig:exp-sensitivity-scaling}
\end{figure}

\noindent\textbf{Multi-GPU Scalability.} As shown in Figure~\ref{fig:exp-sensitivity-scaling}, we evaluate the frame encoding throughput of \ClusterVLM{} under different numbers of GPUs. The results show that the throughput scales nearly linearly with the number of GPUs. Under data partitioning, each GPU performs KVCache clustering independently, incurring no additional communication overhead. This scalability highlights the potential of \ClusterVLM{} for deployment in larger-scale real-world streaming applications.

\section{Related Work}
\noindent\textbf{Long Video Understanding.} Existing VLMs, such as Video-LLaVA~\cite{video-llava} and LLaVA-OneVision~\cite{llava-ov}, typically support uniform frame sampling and token compression. To support longer contexts, prior work~\cite{dyto, aks, adaretake, longvu, moviechat} has explored adaptive keyframe sampling, dynamic token merging, and memory compression. In addition, recent methods such as ViTL~\cite{vitl} introduce interleaved reasoning for span grounding. However, these approaches primarily follow an offline paradigm, assuming that the entire video is available before a query is issued. As a result, they rely on static, offline KVCache processing and are not applicable to infinite, real-time video streams under constrained GPU memory.

\noindent\textbf{Streaming Video Understanding.} Recent studies have begun exploring streaming video processing. VideoLLM-Online~\cite{videollm-online} and StreamingVLM~\cite{streaming-vlm} enable real-time dialogue while maintaining a fixed-size context. Several works~\cite{quick-video, videostreaming, timechat-online} improve streaming efficiency through system co-design or dynamic token dropping. To mitigate the linearly growing memory footprint, training-free KVCache compression methods, such as LiveVLM~\cite{live-vlm} and StreamMem~\cite{stream-mem}, have also been proposed. However, prioritizing online efficiency often comes at the expense of structured KVCache organization. Without well-organized KVCache management, existing approaches inevitably suffer from severe long-term information forgetting, leading to accuracy degradation on complex queries that require reasoning over the entire video history.

\noindent\textbf{VLM KVCache Offloading.} To overcome memory bottlenecks, a promising approach is to offload KVCache to CPU memory and retrieve it dynamically based on attention sparsity. Early work~\cite{long-exposure,jenga,neuralink,dynakv,swarm,infllm} explored this idea for LLMs, and later studies~\cite{rekv,live-vlm,stream-mem} extended in-context KVCache retrieval to streaming video scenarios. These methods effectively reduce both computational and memory overhead. However, all existing approaches organize KVCache at the token level, which incurs substantial system overhead. Meanwhile, several systems work~\cite{hydrogen,coaxial,mcfuser,pipeinfer,mlp-offload} seek to address these inefficiencies by better exploiting hardware capabilities. Building on these insights, \ClusterVLM{} adopts an algorithm-system co-design centered on cross-modal KVCache clustering in VLMs, achieving both higher accuracy and better system performance.

\section{Conclusion}
We present \ClusterVLM{}, a cluster-driven VLM inference system for streaming long video understanding. By leveraging the cross-modal clustering of VLM KVCache, \ClusterVLM{} enables efficient KVCache organization, maintenance, and retrieval, leading to improved overall system performance.

\bibliography{references}

@techreport{gemini3-pro,
    title        = {Gemini 3 Pro Model Card},
    author       = {{Google DeepMind}},
    year         = {2025},
    month        = nov,
    institution  = {Google DeepMind},
    url          = {https://storage.googleapis.com/deepmind-media/Model-Cards/Gemini-3-Pro-Model-Card.pdf},
    note         = {Accessed: December 2025}
}

@article{gpt-5,
    title={Openai gpt-5 system card},
    author={Singh, Aaditya and Fry, Adam and Perelman, Adam and Tart, Adam and Ganesh, Adi and El-Kishky, Ahmed and McLaughlin, Aidan and Low, Aiden and Ostrow, AJ and Ananthram, Akhila and others},
    journal={arXiv preprint arXiv:2601.03267},
    year={2025},
    url = {https://doi.org/10.48550/arXiv.2601.03267}
}

@misc{vit,
    title={An Image is Worth 16x16 Words: Transformers for Image Recognition at Scale}, 
    author={Alexey Dosovitskiy and Lucas Beyer and Alexander Kolesnikov and Dirk Weissenborn and Xiaohua Zhai and Thomas Unterthiner and Mostafa Dehghani and Matthias Minderer and Georg Heigold and Sylvain Gelly and Jakob Uszkoreit and Neil Houlsby},
    year={2021},
    eprint={2010.11929},
    archivePrefix={arXiv},
    primaryClass={cs.CV},
    url = {https://doi.org/10.48550/arXiv.2010.11929}
}

@article{drive-gpt4,
    title={Drivegpt4: Interpretable end-to-end autonomous driving via large language model},
    author={Xu, Zhenhua and Zhang, Yujia and Xie, Enze and Zhao, Zhen and Guo, Yong and Wong, Kwan-Yee K and Li, Zhenguo and Zhao, Hengshuang},
    journal={IEEE Robotics and Automation Letters},
    volume={9},
    number={10},
    pages={8186--8193},
    year={2024},
    publisher={IEEE},
    url = {https://doi.org/10.1109/LRA.2024.3440097}
}

@inproceedings{lm-drive,
    title={Lmdrive: Closed-loop end-to-end driving with large language models},
    author={Shao, Hao and Hu, Yuxuan and Wang, Letian and Song, Guanglu and Waslander, Steven L and Liu, Yu and Li, Hongsheng},
    booktitle={Proceedings of the IEEE/CVF conference on computer vision and pattern recognition},
    pages={15120--15130},
    year={2024},
    url={https://doi.org/10.1109/CVPR52733.2024.01432}
}

@article{palm-e,
    title={Palm-e: An embodied multimodal language model},
    author={Driess, Danny and Xia, Fei and Sajjadi, Mehdi SM and Lynch, Corey and Chowdhery, Aakanksha and Wahid, Ayzaan and Tompson, Jonathan and Vuong, Quan and Yu, Tianhe and Huang, Wenlong and others},
    journal={arXiv preprint arXiv:2303.03378},
    year={2023},
    url={https://doi.org/10.48550/arXiv.2303.03378}
}

@article{open-vla,
    title={Openvla: An open-source vision-language-action model},
    author={Kim, Moo Jin and Pertsch, Karl and Karamcheti, Siddharth and Xiao, Ted and Balakrishna, Ashwin and Nair, Suraj and Rafailov, Rafael and Foster, Ethan and Lam, Grace and Sanketi, Pannag and others},
    journal={arXiv preprint arXiv:2406.09246},
    year={2024},
    url={https://doi.org/10.48550/arXiv.2406.09246}
}

@article{llava,
    title={Visual instruction tuning},
    author={Liu, Haotian and Li, Chunyuan and Wu, Qingyang and Lee, Yong Jae},
    journal={Advances in neural information processing systems},
    volume={36},
    pages={34892--34916},
    year={2023},
    url={https://doi.org/10.48550/arXiv.2304.08485}
}

@article{ferret,
    title={Ferret: Refer and ground anything anywhere at any granularity},
    author={You, Haoxuan and Zhang, Haotian and Gan, Zhe and Du, Xianzhi and Zhang, Bowen and Wang, Zirui and Cao, Liangliang and Chang, Shih-Fu and Yang, Yinfei},
    journal={arXiv preprint arXiv:2310.07704},
    year={2023},
    url={https://doi.org/10.48550/arXiv.2310.07704}
}

@inproceedings{video-llava,
    title={Video-llava: Learning united visual representation by alignment before projection},
    author={Lin, Bin and Ye, Yang and Zhu, Bin and Cui, Jiaxi and Ning, Munan and Jin, Peng and Yuan, Li},
    booktitle={Proceedings of the 2024 conference on empirical methods in natural language processing},
    pages={5971--5984},
    year={2024},
    url={https://aclanthology.org/2024.emnlp-main.342}
}

@article{video-chat,
    title={Videochat: Chat-centric video understanding},
    author={Li, KunChang and He, Yinan and Wang, Yi and Li, Yizhuo and Wang, Wenhai and Luo, Ping and Wang, Yali and Wang, Limin and Qiao, Yu},
    journal={Science China Information Sciences},
    volume={68},
    number={10},
    pages={200102},
    year={2025},
    publisher={Springer},
    url={https://doi.org/10.1007/s11432-024-4321-9}
}

@article{rekv,
    title={Streaming video question-answering with in-context video kv-cache retrieval},
    author={Di, Shangzhe and Yu, Zhelun and Zhang, Guanghao and Li, Haoyuan and Zhong, Tao and Cheng, Hao and Li, Bolin and He, Wanggui and Shu, Fangxun and Jiang, Hao},
    journal={arXiv preprint arXiv:2503.00540},
    year={2025},
    url={https://doi.org/10.48550/arXiv.2503.00540}
}

@article{live-vlm,
    title={Livevlm: Efficient online video understanding via streaming-oriented kv cache and retrieval},
    author={Ning, Zhenyu and Liu, Guangda and Jin, Qihao and Ding, Wenchao and Guo, Minyi and Zhao, Jieru},
    journal={arXiv preprint arXiv:2505.15269},
    year={2025},
    url={https://doi.org/10.48550/arXiv.2505.15269}
}

@article{stream-mem,
    title={Streammem: Query-agnostic kv cache memory for streaming video understanding},
    author={Yang, Yanlai and Zhao, Zhuokai and Shukla, Satya Narayan and Singh, Aashu and Mishra, Shlok Kumar and Zhang, Lizhu and Ren, Mengye},
    journal={arXiv preprint arXiv:2508.15717},
    year={2025},
    url={https://doi.org/10.48550/arXiv.2508.15717}
}

@inproceedings{videollm-online,
    author= {Joya Chen and Zhaoyang Lv and Shiwei Wu and Kevin Qinghong Lin and Chenan Song and Difei Gao and Jia-Wei Liu and Ziteng Gao and Dongxing Mao and Mike Zheng Shou},
    title= {VideoLLM-online: Online Video Large Language Model for Streaming Video},
    booktitle= {CVPR},
    year= {2024},
    url={https://doi.org/10.1109/CVPR52733.2024.01742}
}

@misc{streaming-vlm,
    title={StreamingVLM: Real-Time Understanding for Infinite Video Streams}, 
    author={Ruyi Xu and Guangxuan Xiao and Yukang Chen and Liuning He and Kelly Peng and Yao Lu and Song Han},
    year={2025},
    eprint={2510.09608},
    archivePrefix={arXiv},
    primaryClass={cs.CV},
    url={https://doi.org/10.48550/arXiv.2510.09608}, 
}

@article{quick-video,
    title={QuickVideo: Real-Time Long Video Understanding with System Algorithm Co-Design},
    author={Schneider, Benjamin and Jiang, Dongfu and Du, Chao and Pang, Tianyu and Chen, Wenhu},
    journal={arXiv preprint arXiv:2505.16175},
    year={2025},
    eprint={2505.16175},
    archivePrefix={arXiv},
    primaryClass={cs.CV},
    url={https://doi.org/10.48550/arXiv.2505.16175}
}

@article{vitl,
    title={Video-in-the-Loop: Span-Grounded Long Video QA with Interleaved Reasoning},
    author={Wang, Chendong and Bai, Donglin and Yang, Yifan and Jin, Xiao and Zhang, Anlan and Wang, Rui and Jiang, Shiqi and Yang, Yuqing and Wu, Hao and Dai, Qi and Luo, Chong and Cao, Ting and Qiu, Lili and Banerjee, Suman},
    journal={arXiv preprint arXiv:2510.04022},
    year={2025},
    eprint={2510.04022},
    archivePrefix={arXiv},
    primaryClass={cs.CV},
    url={https://doi.org/10.48550/arXiv.2510.04022}
}

@InProceedings{resnet,
    author = {He, Kaiming and Zhang, Xiangyu and Ren, Shaoqing and Sun, Jian},
    title = {Deep Residual Learning for Image Recognition},
    booktitle = {Proceedings of the IEEE Conference on Computer Vision and Pattern Recognition (CVPR)},
    month = {June},
    year = {2016},
    url={https://doi.org/10.1109/CVPR.2016.90}
}

@article{kmeans,
    title={Least squares quantization in PCM},
    author={Lloyd, Stuart},
    journal={IEEE transactions on information theory},
    volume={28},
    number={2},
    pages={129--137},
    year={1982},
    publisher={IEEE},
    url={https://doi.org/10.1109/TIT.1982.1056489}
}

@misc{llava-ov,
    title={LLaVA-OneVision: Easy Visual Task Transfer}, 
    author={Bo Li and Yuanhan Zhang and Dong Guo and Renrui Zhang and Feng Li and Hao Zhang and Kaichen Zhang and Peiyuan Zhang and Yanwei Li and Ziwei Liu and Chunyuan Li},
    year={2024},
    eprint={2408.03326},
    archivePrefix={arXiv},
    primaryClass={cs.CV},
    url={https://doi.org/10.48550/arXiv.2408.03326}, 
}

@misc{qwen25-vl,
    title={Qwen2.5-VL Technical Report}, 
    author={Shuai Bai and Keqin Chen and Xuejing Liu and Jialin Wang and Wenbin Ge and Sibo Song and Kai Dang and Peng Wang and Shijie Wang and Jun Tang and Humen Zhong and Yuanzhi Zhu and Mingkun Yang and Zhaohai Li and Jianqiang Wan and Pengfei Wang and Wei Ding and Zheren Fu and Yiheng Xu and Jiabo Ye and Xi Zhang and Tianbao Xie and Zesen Cheng and Hang Zhang and Zhibo Yang and Haiyang Xu and Junyang Lin},
    year={2025},
    eprint={2502.13923},
    archivePrefix={arXiv},
    primaryClass={cs.CV},
    url={https://doi.org/10.48550/arXiv.2502.13923}, 
}

@misc{kuaishou,
    author       = {{Kuaishou Technology}},
    title        = {Kuaishou},
    year         = {2026},
    howpublished = {\url{https://www.kuaishou.com/en}},
    note         = {Accessed: 2026-04-08}
}

@misc{mlvu,
    title={MLVU: Benchmarking Multi-task Long Video Understanding}, 
    author={Junjie Zhou and Yan Shu and Bo Zhao and Boya Wu and Zhengyang Liang and Shitao Xiao and Minghao Qin and Xi Yang and Yongping Xiong and Bo Zhang and Tiejun Huang and Zheng Liu},
    year={2025},
    eprint={2406.04264},
    archivePrefix={arXiv},
    primaryClass={cs.CV},
    url={https://doi.org/10.48550/arXiv.2406.04264}, 
}

@misc{long-video-bench,
    title={LongVideoBench: A Benchmark for Long-context Interleaved Video-Language Understanding}, 
    author={Haoning Wu and Dongxu Li and Bei Chen and Junnan Li},
    year={2024},
    eprint={2407.15754},
    archivePrefix={arXiv},
    primaryClass={cs.CV},
    url={https://doi.org/10.48550/arXiv.2407.15754}, 
}

@misc{video-mme,
    title={Video-MME: The First-Ever Comprehensive Evaluation Benchmark of Multi-modal LLMs in Video Analysis}, 
    author={Chaoyou Fu and Yuhan Dai and Yongdong Luo and Lei Li and Shuhuai Ren and Renrui Zhang and Zihan Wang and Chenyu Zhou and Yunhang Shen and Mengdan Zhang and Peixian Chen and Yanwei Li and Shaohui Lin and Sirui Zhao and Ke Li and Tong Xu and Xiawu Zheng and Enhong Chen and Caifeng Shan and Ran He and Xing Sun},
    year={2025},
    eprint={2405.21075},
    archivePrefix={arXiv},
    primaryClass={cs.CV},
    url={https://doi.org/10.48550/arXiv.2405.21075}, 
}

@misc{rvs,
    title={Flash-VStream: Memory-Based Real-Time Understanding for Long Video Streams}, 
    author={Haoji Zhang and Yiqin Wang and Yansong Tang and Yong Liu and Jiashi Feng and Jifeng Dai and Xiaojie Jin},
    year={2024},
    eprint={2406.08085},
    archivePrefix={arXiv},
    primaryClass={cs.CV},
    url={https://doi.org/10.48550/arXiv.2506.23825}, 
}

@InProceedings{dyto,
    author    = {Zhang, Yiming and Zhao, Zhuokai and Chen, Zhaorun and Ding, Zenghui and Yang, Xianjun and Sun, Yining},
    title     = {Beyond Training: Dynamic Token Merging for Zero-Shot Video Understanding},
    booktitle = {Proceedings of the IEEE/CVF International Conference on Computer Vision (ICCV)},
    month     = {October},
    year      = {2025},
    pages     = {22046-22055},
    url={https://doi.org/10.48550/arXiv.2411.14401}
}

@article{aks,
    title={Adaptive Keyframe Sampling for Long Video Understanding},
    author={Tang, Xi and Qiu, Jihao and Xie, Lingxi and Tian, Yunjie and Jiao, Jianbin and Ye, Qixiang},
    journal={arXiv preprint arXiv:2502.21271},
    year={2025},
    url={https://doi.org/10.1109/CVPR52734.2025.02711}
}

@article{adaretake,
    title={AdaReTaKe: Adaptive Redundancy Reduction to Perceive Longer for Video-language Understanding},
    author={Wang, Xiao and Si, Qingyi and Wu, Jianlong and Zhu, Shiyu and Cao, Li and Nie, Liqiang},
    journal={arXiv preprint arXiv:2503.12559},
    year={2025},
    url={https://doi.org/10.48550/arXiv.2503.12559}
}

@article{longvu,
    author = {Shen, Xiaoqian and Xiong, Yunyang and Zhao, Changsheng and Wu, Lemeng and Chen, Jun and Zhu, Chenchen and Liu, Zechun and Xiao, Fanyi and Varadarajan, Balakrishnan and Bordes, Florian and Liu, Zhuang and Xu, Hu and Kim, Hyunwoo J. and Soran, Bilge and Krishnamoorthi, Raghuraman and Elhoseiny, Mohamed and Chandra, Vikas}, 
    title = {LongVU: Spatiotemporal Adaptive Compression for Long Video-Language Understanding},
    journal = {arXiv preprint arXiv:2410.17434},
    year = {2024},
    url={https://doi.org/10.48550/arXiv.2410.17434}
}

@article{moviechat,
    title={MovieChat: From Dense Token to Sparse Memory for Long Video Understanding},
    author={Song, Enxin and Chai, Wenhao and Wang, Guanhong and Zhang, Yucheng and Zhou, Haoyang and Wu, Feiyang and Guo, Xun and Ye, Tian and Lu, Yan and Hwang, Jenq-Neng and others},
    journal={arXiv preprint arXiv:2307.16449},
    year={2023},
    url={https://doi.org/10.1109/CVPR52733.2024.01725}
}

@article{videostreaming,
    title={Streaming Long Video Understanding with Large Language Models},
    author={Qian, Rui and Dong, Xiaoyi and Zhang, Pan and Zang, Yuhang and Ding, Shuangrui and Lin, Dahua and Wang, Jiaqi},
    journal={https://arxiv.org/abs/2405.16009},
    year={2024},
    url={https://doi.org/10.52202/079017-3792}
}

@article{timechat-online,
    title={TimeChat-Online: 80\% Visual Tokens are Naturally Redundant in Streaming Videos},
    author={Yao, Linli and Li, Yicheng and Wei, Yuancheng and Li, Lei and Ren, Shuhuai and Liu, Yuanxin and Ouyang, Kun and Wang, Lean and Li, Shicheng and Li, Sida and Kong, Lingpeng and Liu, Qi and Zhang, Yuanxing and Sun, Xu},
    journal={https://arxiv.org/abs/2504.17343},
    year={2025},
    url={https://doi.org/10.1145/3746027.3754839}
}

@article{infllm,
    title={InfLLM: Training-Free Long-Context Extrapolation for LLMs with an Efficient Context Memory},
    author={Xiao, Chaojun and Zhang, Pengle and Han, Xu and Xiao, Guangxuan and Lin, Yankai and Zhang, Zhengyan and Liu, Zhiyuan and Sun, Maosong},
    journal={https://arxiv.org/abs/2402.04617},
    year={2024},
    url={https://doi.org/10.52202/079017-3801}
}

@INPROCEEDINGS{hydrogen,
    title={Hydrogen: Contention-Aware Hybrid Memory for Heterogeneous CPU-GPU Architectures},
    author={Li, Yiwei and Gao, Mingyu},
    booktitle={SC24: International Conference for High Performance Computing, Networking, Storage and Analysis}, 
    year={2024},
    url={https://doi.org/10.1109/SC41406.2024.00017}
}

@inproceedings{coaxial,
    author = {Cho, Albert and Saxena, Anish and Qureshi, Moinuddin and Daglis, Alexandros},
    title = {COAXIAL: A CXL-Centric Memory System for Scalable Servers},
    year = {2024},
    booktitle = {Proceedings of the International Conference for High Performance Computing, Networking, Storage, and Analysis},
    url = {https://doi.org/10.1109/SC41406.2024.00101}
}

@inproceedings{mcfuser,
    author = {Zhang, Zheng and Yang, Donglin and Zhou, Xiaobo and Cheng, Dazhao},
    title = {MCFuser: High-Performance and Rapid Fusion of Memory-Bound Compute-Intensive Operators},
    year = {2024},
    url = {https://doi.org/10.1109/SC41406.2024.00040},
    booktitle = {Proceedings of the International Conference for High Performance Computing, Networking, Storage, and Analysis}
}

@INPROCEEDINGS{pipeinfer,
  author={Butler, Branden and Yu, Sixing and Mazaheri, Arya and Jannesari, Ali},
  booktitle={SC24: International Conference for High Performance Computing, Networking, Storage and Analysis}, 
  title={PipeInfer: Accelerating LLM Inference using Asynchronous Pipelined Speculation}, 
  year={2024},
  url = {https://doi.org/10.1109/SC41406.2024.00046}
}

@inproceedings{mlp-offload,
    author = {Maurya, Avinash Kumar and Rafique, M. Mustafa and Cappello, Franck and Nicolae, Bogdan},
    title = {MLP-Offload: Multi-Level, Multi-Path Offloading for LLM Pre-training to Break the GPU Memory Wall},
    year = {2025},
    url = {https://doi.org/10.1145/3712285.3759864},
    booktitle = {Proceedings of the International Conference for High Performance Computing, Networking, Storage and Analysis}
}

@inproceedings{long-exposure,
  title={LONG EXPOSURE: Accelerating Parameter-Efficient Fine-Tuning for LLMs under Shadowy Sparsity},
  author={Wang, Tuowei and Li, Kun and Hao, Zixu and Bai, Donglin and Ren, Ju and Zhang, Yaoxue and Cao, Ting and Yang, Mao},
  booktitle={SC24: International Conference for High Performance Computing, Networking, Storage and Analysis},
  pages={1--18},
  year={2024},
  organization={IEEE},
  url={https://doi.org/10.1109/SC41406.2024.00081}
}

@inproceedings{jenga,
  title={$\{$JENGA$\}$: Enhancing $\{$LLM$\}$$\{$Long-Context$\}$ Fine-tuning with Contextual Token Sparsity},
  author={Wang, Tuowei and Chen, Xingyu and Li, Kun and Cao, Ting and Ren, Ju and Zhang, Yaoxue},
  booktitle={2025 USENIX Annual Technical Conference (USENIX ATC 25)},
  pages={123--141},
  year={2025},
  url={https://doi.org/10.48550/arXiv.2501.09767}
}

@inproceedings{neuralink,
  title={Neuralink: Fast on-Device LLM Inference with Neuron Co-Activation Linking},
  author={Wang, Tuowei and Fan, Ruwen and Huang, Minxing and Hao, Zixu and Li, Kun and Cao, Ting and Lu, Youyou and Zhang, Yaoxue and Ren, Ju},
  booktitle={Proceedings of the 30th ACM International Conference on Architectural Support for Programming Languages and Operating Systems, Volume 3},
  pages={147--162},
  year={2025},
  url={https://doi.org/10.1145/3676642.3736114}
}

@misc{dynakv,
      title={DynaKV: Enabling Accurate and Efficient Long-Sequence LLM Decoding on Smartphones}, 
      author={Tuowei Wang and Minxing Huang and Fengzu Li and Ligeng Chen and Jinrui Zhang and Ju Ren},
      year={2025},
      eprint={2511.07427},
      archivePrefix={arXiv},
      primaryClass={cs.DC},
      url={https://doi.org/10.48550/arXiv.2511.07427}, 
}

@misc{swarm,
      title={Swarm: Co-Activation Aware KVCache Offloading Across Multiple SSDs}, 
      author={Tuowei Wang and Liyun Chu and Ruwen Fan and Ju Ren},
      year={2026},
      eprint={2603.17803},
      archivePrefix={arXiv},
      primaryClass={cs.PF},
      url={https://doi.org/10.48550/arXiv.2603.17803}, 
}
\bibliographystyle{IEEEtran}

\end{document}